\documentclass[11pt, a4paper]{article}
\usepackage[colorlinks=true, pdfstartview=FitV, linkcolor=blue, citecolor=blue, urlcolor=blue]{hyperref}
\usepackage{amsmath,amssymb}
\usepackage{graphicx}
\usepackage{cite}

\newcommand{\tr}{\text{tr}}
\newcommand{\Nc}{N_{\text{c}}}
\newcommand{\Nf}{N_{\text{f}}}
\newcommand{\Lag}{\mathcal{L}}
\newcommand{\calP}{$\mathcal{P}$ }
\newcommand{\calCP}{$\mathcal{CP}$ }
\newcommand{\LQCD}{\Lambda_{\text{QCD}}}

\begin{document}
\title{ {\bf QCD $\theta$-vacua from the chiral limit \\to the quenched limit} }
\author{Kazuya Mameda \\
\\ \vspace{6pt} {\it Department of Physics, The University of Tokyo, Hongo, Tokyo 113-0031, Japan} }
\date{ }
\maketitle
\begin{abstract}
We investigate the dependence of the QCD vacuum structure on the $\theta$-angle and quark mass, using the Di-Vecchia--Veneziano model. Although the Di-Vecchia--Veneziano model is a chiral effective model, it contains the topological properties of the pure Yang--Mills theory. It is shown that within this model, the ground state energies for all $\theta$ are continuous functions of quark mass from the chiral limit to the quenched limit, even including the first order phase transition at $\theta = \pi$. Based on this effective model, we discuss (i) how the ground state depends on quark mass, and (ii) why the phase transition at $\theta = \pi$ is present both in the chiral and quenched limit. In order to analyze the relation between quark mass and the $\theta$-vacua, we calculate the chiral condensate as a function of quark mass. Also, considering the presence of the innate metastable states included in the QCD $\theta$-vacuum, we also give a unified understanding of the phase transitions at $\theta = \pi$ in the chiral and quenched limit.
\end{abstract}

\section{Introduction}\label{sec:intro}
An SU(N) gauge theory contains the topology which originates in topologically non-trivial excitations of the gauge field. The vacuum structure in Quantum Chromodynamics (QCD) is directly influenced from such a topology because QCD vacuum is defined by the parameter, so-called $\theta$-angle, which enters in the QCD action as the Chern--Simons term $i Q \theta$ (where $Q$ denotes the winding number of gluon, $Q = \int d^4 x \frac{g^2}{32\pi} F^a_{\mu\nu} \tilde F_a^{\mu\nu}$). In principle, the actual value of the $\theta$-angle is so important for investigations of QCD phenomena. (Note that any transition between different $\theta$-vacua is forbidden in infinite volume.) The \calCP symmetry in QCD, especially, depends only on the value of the $\theta$-angle since the Chern--Simons term violates \calCP. The $\theta = 0$ case is the most reliable candidate maintained by an experimental result $\theta \leq 10^{-10}$~\cite {Baker:2006ts} and in fact, phenomena in terms of the strong interaction seem not to break \calCP symmetry. Unfortunately, we do not know the theoretical reason why $\theta = 0$ is selected. This is one of the unsolved problems in the standard model, referred to as the strong \calCP problem~\cite {Cheng:1987gp, Peccei:1998jt}. 

One might wonder if it is necessary to study the topological property in QCD, because of the fact that the $\theta$-angle is found to have a negligibly small value. Nevertheless, we should consider the $\theta$-vacuum structure for investigations of QCD since there exists another origin of the topology in QCD, i.e. the axial anomaly. For example, the $\eta '$-meson acquires, through the anomaly, an extra mass much larger than the mass of other mesons which seem to be the pseudo Nambu--Goldstone bosons related to the chiral symmetry breaking~\cite{Witten:1979vv,Veneziano:1979ec}. Lattice QCD simulations confirm that the $\eta'$ mass formula gives a value consistent with the observed one. We have to note here that the mass formula suggested in Ref.~\cite{Witten:1979vv,Veneziano:1979ec} includes only quantities at $\theta = 0$. It is indicated, hence, that even if $\theta = 0$, the topological nature of QCD can give contributions to physics. (It has been also proposed recently that dark energy could stem from the $\theta$ -vacuum at $\theta = 0$~\cite{Urban:2009vy,Urban:2009yg,Ohta:2010in}.)

Among the physics involved with the QCD topology, phenomena with $\theta \neq 0$ are on the frontier of the most important subjects in QCD. It has been suggested that the $\theta$-angle can take various values event by event in heavy ion collisions~\cite{Morley:1983wr,Kharzeev:1998kz}. In other words, the $\theta$-angle in the collision is considered not as a constant parameter, but a function of space-time coordinates (or like axion, which is the dynamical field introduced to solve the strong \calCP problem~\cite {Peccei:1977hh}). Such a new possibility of the $\theta$-angle leads to interesting phenomena related to the topological structure of QCD~\cite {Vilenkin:1980fu, Son:2004tq, Metlitski:2005pr, Kharzeev:2007jp, Fukushima:2008xe,Kharzeev:2010gd}.

The presence of an experimentally accessible situation with a finite $\theta$-angle gives a strong motivation for the study of the $\theta$ -vacuum. Specifically, there is a need to investigate what happens when $\theta = 0$ changes into finite values. Indeed for the structure for a finite $\theta$-angle, the vacuum structure has been the subject of detailed studies over the years, and it has been shown that the case where $\theta = \pi$ is of particular importance~\cite{Dashen:1970et, Coleman:1976uz,Witten:1980sp,'tHooft:1981ht,Ohta:1981ai,Wiese:1988qz, Affleck:1991tj,Creutz:1994px,Creutz:1995wf,Creutz:2009kx, Creutz:2010ee, Hosotani:1996hy,Smilga:1998dh,Witten:1998uka, Halperin:1998rc,Boer:2008ct,Boomsma:2009eh,D'Elia:2012vv,D'Elia:2013eua,Aoki:2014moa}. (See also Ref.~\cite{Vicari:2008jw}, which summarizes works related to the $\theta$-dependence of QCD.) Naively, it might seem that $\theta = \pi$ is merely another candidate where QCD is \calCP invariant, and that there is nothing particularly special about $\theta = \pi$. Nevertheless, $\theta = \pi$ does give rise to intriguing physics: the first order phase transition that leads to spontaneous \calCP symmetry breaking. This is one of the main points discussed in this paper.

As we have already mentioned, the axial symmetry is deeply connected to the QCD topology, as well as non-trivial gluonic excitations. We can assume, then, that the quark sector is strongly affected by this structure originating from the Chern--Simons term. Namely, by an axial rotation of the fermion field, the $\theta$-dependence moves to the quark mass term from the Chern--Simons term. This implies that the $\theta$-vacua depend not only on the strong $\theta$-angle but also on quark mass. Here, the best quantity which explicitly shows the mass-dependence of the $\theta$-vacuum is the topological susceptibility, $\chi_{\rm top}(m)$, interpreted as a function of quark mass. Indeed the topological susceptibilities in the chiral limit and the quenched limit have quite different forms as a function of the mass. It is important, hence, that we investigate the mass-dependence of the $\theta$-vacuum, in order to completely understand the topological nature in QCD. Nevertheless, the vacuum structure for arbitrary quark mass has not been studied in detail, except for the chiral theory and the pure gluonic theory. (See Ref.~\cite{Tytgat:1999yx}, where differences between the chiral and quenched limit are argued, comparing the cases that $\Nf = 0$ and that $\Nf \neq 0$.) Briefly speaking, we do not know the vacuum structure for quarks heavier than $\LQCD$, and no effective theory including the $\theta$-angle is known. (In Ref.\cite{Soto:1992vn}, for example, it is pointed out that effects of the axial anomaly definitely exist in the heavy quark effective theory, but not method for calculating these contributions to dynamics is proposed.)

The Di-Vecchia--Veneziano model~\cite{DiVecchia:1980ve} is one of the simplest, yet remarkable chiral effective models reflecting the topological properties of QCD, and thus the dynamics described by this model have been investigated from many points of view~\cite{Kawarabayashi:1980dp,Leutwyler:1992yt, Kharzeev:1998kz, Blaschke:2001ek, Urban:2009yg}. (It is also possible to discuss a phenomenon related to $\theta$ and electromagnetic field $\boldsymbol B$, based on the Di-Vecchia--Veneziano model~\cite{Fukushima:2012fg}.) In this paper, we discuss the $\theta$-vacuum structure from the point of view of the mass-dependence, within the Di-Vecchia--Veneziano model. In Sec.~\ref{sec:mdth}, we mention the relation between quark mass and the topology more specifically. In Sec.~\ref{sec:VD}, we show that this model is not a mere chiral model containing the QCD topology, but a more remarkable model that includes even the features of the $\theta$-vacua in the quenched limit. In other words, the suggestion from this chiral model is quite pedagogical even for large mass. Based on this idea, in Sec.~\ref{sec:anly}, we calculate the vacuum energies for arbitrary quark mass. In Sec.~\ref{sec:cc}, by calculating chiral condensate as a function of quark mass within this model, we reveal the dynamical mechanism of the $\theta$-vacuum, specifically the role of quark mass in the decision of vacuum states. Moreover, we investigate the first order phase transition at $\theta = \pi$, which is quite beneficial for the understanding of the vacuum structure as a function of quark mass. It is important, here, that the phase transition in the chiral limit and the one in the quenched limit seem naively similar, although the dynamics in two limits are completely different. In Sec.~\ref{sec:ms}, we discuss the origin of such a similarity of the vacuum structure, on the basis of the degeneracy of the metastable vacua included in QCD vacuum.

\section{Mass-dependence of the $\theta$-vacuum} \label{sec:mdth} 
Quark mass is one of the most important factors when discussing topological properties of QCD. Let us review the global symmetries in QCD and the axial anomaly here, in order to reveal the mass-dependence of the $\theta$-vacuum. The classical Lagrangian of massless QCD with $\Nf$ flavors is invariant under the transformations related to the global symmetry group $\mathrm{SU({\Nf})_L \times SU({\Nf})_R \times U(1)_V \times U(1)_A}$. In contrast, in quantum theory, the axial symmetry ${\rm U(1)_A}$ is broken due to the axial anomaly. Specifically, the measure of the functional integration of the quark field changes under the ${\rm U(1)_A}$-rotation with a phase $\alpha$ as ${\cal D}q \to {\cal D}q \exp ({ - i \Nf \alpha \omega} )$, where $\omega$ denotes the winding number density i.e., $\int d^4x \omega = Q$. The measure of anti-quark field is transformed similarly. Thus, by this rotation, an extra term is added to the Lagrangian; ${\cal L} \to {\cal L} - 2\Nf \alpha \omega$. Choosing the rotation angle as $\alpha = \theta / 2\Nf$, one can cancel the Chern--Simons term $iQ\theta$. In massive QCD, however, the $\theta$-angle remains in the Lagrangian even though under the same rotation, due to the extra factor appearing the quark mass term, which is not invariant under the ${\rm U(1)_A}$-rotation; $m \to m e^{i\gamma_5 \theta /\Nf}$. Hence, one can schematically write this transformation of the QCD action, as follows;
\begin{equation}\label{eq:axial}
 S_\theta = S (m) + iQ\theta\ \overset{{\rm Axial}}{\longrightarrow} 
\ S_\theta = S (me^{i\gamma_5 \theta /\Nf}) . 
\end{equation}
This means that the $\theta$-dependence of QCD enters only through the factor coupled with quark mass, and thus the $\theta$-vacuum structure depends not only on $\theta$ but also on $m$.

The mass-dependence of the QCD $\theta$-vacua emerges in typical limits: the chiral limit and the quenched limit. In these two limits, the topological susceptibility
\begin{equation}
 \chi_{\rm top} = \frac{1}{V_4} \int d^4 x \langle \omega (x) \omega (0) \rangle \Big|_{\theta = 0},
\end{equation}
for example, takes completely different values. Lattice QCD simulation shows that the susceptibility in the pure Yang--Mills theory is a constant; $\chi_{\rm top} = \chi_{\rm pure} \sim (170-180{\rm MeV})^4$~\cite{Alles:1996nm}. On the other hand, it is derived that for small quark mass, the topological susceptibility is proportional to the mass; $\chi_{\rm top}\propto m$~\cite{Leutwyler:1992yt}. (This is consistent with the fact that the $\theta$-dependence through the mass term vanishes in the chiral limit, as shown in Eq.~\eqref{eq:axial}.) Since the topological susceptibility is the potential curvature of the $\theta$-vacuum at $\theta = 0$, the difference of the topological susceptibilities in two limits actually indicates that the vacuum structure depends on quark mass.

For middle mass, i.e., finite large mass, the $\theta$-vacuum structure is, however, not clear. This is due to the fact that neither a heavy quark effective model or theory including the $\theta$-angle is known, and thus there is no method for calculating the physical quantities under the presence of the $\theta$-angle for heavy quarks, compared with $\LQCD$ (i.e., $c$-, $t$-, and $b$-quarks). The unclarity of the $\theta$-vacuum for large mass might be realized if one reconsiders the $\theta$-vacuum in the pure Yang Mills theory, comparing with the one in the heavy quark theory.

In general, the degree of freedom in a theory is a dynamical field. Heavy quarks do not, therefore, engaged in dynamics, or the propagations of heavy flavors become negligible as the masses are increased. This can be rephrased in the language of field theory as follows; the pure Yang--Mills theory is regarded as the theory after one integrates out heavy flavors. From a naive calculation, however, this procedure to obtain the pure Yang--Mills theory from QCD would lead to an incorrect $\theta$-dependence. By the ${\rm U(1)_A}$-rotation, $q \to e^{i\gamma_5 \theta / 2\Nf} q$, as mentioned already, the Chern--Simons term can be canceled, and instead of this, the $\theta$-angle appears only in the mass term. If starting from this axial rotated action, one could integrate out heavy quarks, then $\theta$-angle also have been removed from the action. However, this $\theta$-independent pure Yang--Mills theory lies in direct contradiction to the fact that we can obtain the pure Yang--Mills theory with the definite $\theta$-dependence, starting from the action before the rotation. In fact, as mentioned, the topological susceptibility in the pure Yang--Mills theory has a finite value $\chi_{\rm pure}$.

This inconsistency might be caused by the non-commutativity of the large mass limit and the regularization of the axial anomaly (which is indeed the case that we shall see the cause of the non-commutativity for the vacuum state in a later section). This implies, however, that the heavy quark approximation with the $\theta$-angle is never trivial and the pure Yang--Mills theory cannot be derived straightforwardly from full QCD. Therefore, we should take care the quenched theory in the presence of $\theta$-angle. Additionally the non-commutativity imposes us to analyze more particularly the $\theta$-vacuum structure depending on quark mass. We discuss the dynamical effect of quark mass in Sec.~\ref{sec:cc}, by the estimation of the chiral condensate as a function of quark mass.

The physics at $\theta = \pi$ is one of the most intriguing points in the investigation of the $\theta$-vacua, and also give a good example that demonstrates the mass-dependence. As mentioned in Sec.~\ref{sec:intro}, $\theta = \pi$ is related to the important feature of the $\theta$-vacuum, namely the first order phase transition~\cite{Dashen:1970et}. Using, the Di-Vecchia--Veneziano model, which is a chiral effective model involving the ${\rm U(1)_A}$ anomaly in full QCD, one can see that the phase transition occurs at $\theta = \pi$~\cite{Witten:1980sp}. It is also suggested that the first order phase transition occurs in the quenched theory, i.e., the pure Yang--Mills theory by the investigations with the holographic approach~\cite{Witten:1998uka}. These phase transitions, one in the chiral theory and the other in the pure Yang--Mills theory, might appear as an inherent feature involving in the $\theta$-vacuum. We should note, however, that this similarity is never trivial, and indeed that these phase transitions have completely different underlying mechanics. The phase transition at $\theta =\pi$ in the chiral effective model results from the dynamics of meson fields or quarks. On the other hand, in the pure Yang--Mills theory, of course, there are only gluons which could trigger such a transition. Additionally, the formations of metastable states which leads to the phase transitions are definitely different in the chiral and quenched limits. While in the quenched limit, there are infinite non-degenerate metastable states, those in the chiral limit are categorized into only few patterns. This difference with respect to the metastable states implies, hence, that the degeneracy originates in the number of dynamical flavors. We discuss in Sec.~\ref{sec:ms}, including the point of view of the relation between dynamical flavors and metastable states, the difference of the first order phase transitions in the chiral and quenched limit, based on a unified framework.

\section{Di-Vecchia--Veneziano model for large mass}\label{sec:VD}
The main goal of this paper is to find the the ground state as a function of the $\theta$-angle and quark mass, based on the Di-Vecchia--Veneziano model~\cite{DiVecchia:1980ve} given by the following Lagrangian,
\begin{equation}\label{eq:Lchi}
 \begin{split}
 \Lag = \frac{f_\pi^2}{4} \tr\ \bigl[ \partial_\mu U^\dagger
 \partial^\mu U + 2\chi (MU^\dagger + UM ) \bigr] -\frac{\chi_{\text{pure}}}{2}
 \Bigl[\theta-\frac{i}{2}\tr(\ln U - \ln U^\dagger)
 \Bigr]^2,
 \end{split}
\end{equation}
where $f_\pi$ is the pion decay constant, $\chi = -\langle \bar q q \rangle / f^2_\pi$ is a parameter given by the Gell-Mann--Oakes--Renner relation, and $\chi_{\rm pure}$ stands for the topological susceptibility in the pure Yang--Mills theory. The first trace part are the first order terms of the chiral perturbation theory and the second part, including the $\theta$-angle, reflects the axial anomaly. From this effective Lagrangian, we calculate the $\theta$-vacuum for arbitrary quark mass, following the argument in Ref.~\cite{Witten:1980sp}. Since the Di-Vecchia--Veneziano model is based on the chiral perturbation theory, however, only the exact ground state for small mass is obtained, and indeed in the original work~\cite{Witten:1980sp}, quark masses are assumed as realistic small values. Therefore it is necessary to discuss the validity of the extending this model to the case for heavy quark. 

\begin{figure}
\begin{center}
 \includegraphics[width=0.55\columnwidth]{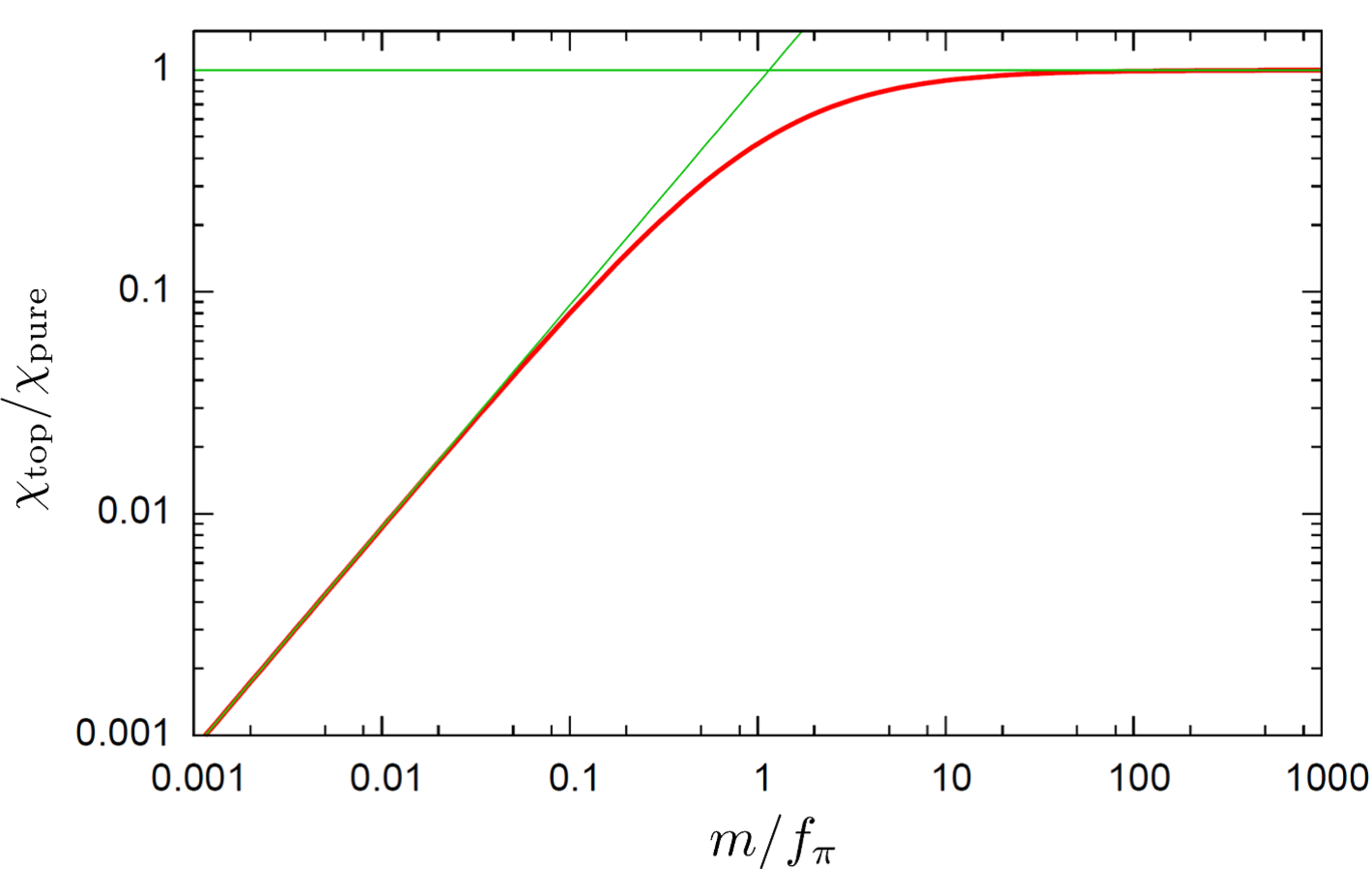}
 \caption{
The topological susceptibility as a function of quark mass where $f_\pi^2/\chi = -\langle\bar{q}q\rangle = (250\ \text{MeV})^3$. For small mass the topological susceptibility is a linear function of quark mass. On the other hand, the topological susceptibility for large mass is asymptotically equal to $\chi_{\rm pure }$. The two green lines in this figure denote $\chi_{\rm top}(m) = m f_\pi ^ 2 \chi / 2$ and $\chi_{\rm top}(m) = \chi _{\rm pure}$, respectively. The Di-Vecchia--Veneziano model, therefore, includes the vacuum structure both in the chiral and quenched limit.
}
 \label{fig:chitop}
\end{center}
\end{figure}

For this purpose, it is useful to analyze the topological susceptibility calculated from this model. For the case of the two degenerate light flavors, namely $M =$ diag $(m,m)$, for example, the topological susceptibility as a function of quark mass is written as~\cite{Leutwyler:1992yt};
\begin{equation}\label{eq:chitop}
\chi_\text{top} (m) = \frac{\chi_\text{pure}m}{m+2\chi_\text{pure}/f_\pi^2\chi}.
\end{equation}
The mass-dependence of the topological susceptibility is indicated in Fig.~\ref{fig:chitop}. Originally, this relation is derived, based on the discussion in ~Ref.~\cite{Witten:1980sp}, where the quark masses are light. However, the derivation of Eq.~\eqref{eq:chitop} does not necessarily require that quark masses are light. If Eq.~\eqref{eq:chitop} is regarded as a valid relation even for large quark mass, then one finds that in the large mass limit, this function approaches to the expected topological susceptibility $\chi_{\rm pure}$. This implies, hence, that the Di-Vecchia--Veneziano model contains the property of the $\theta$-vacuum even in the quenched limit.
 
It is not accidental that the Di-Vecchia--Veneziano model reproduces the result in the pure Yang--Mills theory. The topological property in this model is introduced through the $1/\Nc$ approximation~\cite{DiVecchia:1980ve}. In the large $\Nc$ limit, quarks are decoupled completely from gluons, and thus there are no dynamical quarks after integrating out fermion fields. This situation is equivalent to the one in the large mass limit. The topological property in the large mass limit, therefore, enters automatically in the Di-Vecchia--Veneziano model.

Since higher order mass terms should be considered for large quark mass, the Di-Vecchia--Veneziano model, which contains only the leading order of mass term, could not be used. Nevertheless, the topological susceptibility Eq.~\eqref{eq:chitop} reproduces the result in the pure Yang--Mills theory. In addition to this, in Sec.~\ref{sec:anly}, we will find that the $\theta$-vacuum state calculated with the Di-Vecchia--Veneziano model in the large mass limit becomes exactly the same form as the one obtained in the pure Yang--Mills theory. In this sense, the Di-Vecchia--Veneziano model can be regarded as the simplest model which satisfies both boundary conditions: the topological properties in the chiral limit and the quenched limit. Then we emphasize that the Di-Vecchia--Veneziano model is quite pedagogical for the investigation of the $\theta$-vacuum structure, and at least, the vacuum for sufficiently small and large mass can be obtained correctly from this effective model.

\section{Analysis of the model}\label{sec:anly}
In this section, we consider two quarks with degenerate masses i.e., $M = {\rm diag} (m,m)$. For the ground state energy, the dynamical field should be written as $U={\rm diag}(e^{i\phi_u},e^{i \phi_d})$. The potential part of Eq.~\eqref{eq:Lchi} can then be reduced to the following simpler form;
\begin{equation}\label{eq:potential}
 \begin{split}
V(\phi_u,\phi_d ) = -f_\pi ^2 \chi m(\cos \phi_u + \cos \phi_d ) 
+ \frac{\chi_\mathrm{pure}}{2}(\theta + \phi_u +\phi_d)^2 ,
 \end{split}
\end{equation}
which leads to the set of equations of motion 
\begin{equation}\label{eq:minimize}
f_\pi ^2 \chi m \sin \phi_i + \chi_\mathrm{pure} 
\left(\theta + \phi_u +\phi_d \right) =0, \quad\quad i = u,d.
\end{equation}
The set of solutions which could give the ground state are $\phi_u = \phi_d - 2\pi k= \phi$, where $k$ is an integer since Eq.~\eqref{eq:minimize} implies $\sin \phi_u =\sin \phi_d$. (Another type of solution, $\phi_u = (2k-1)\pi - \phi_d = \phi$ never gives the minimum energy.) Then the equation to solve becomes
\begin{equation}\label{eq:eqk}
f_\pi ^2 \chi m \sin \phi + \chi_\mathrm{pure} 
\left(\theta + 2\pi k + 2\phi \right) =0.
\end{equation}
Especially, in the chiral and quenched limit, this transcendental equation can be solved analytically, and the solutions read
\begin{equation}\label{eq:solution}
\phi (\theta, m) = 
\begin{cases}
{\displaystyle -\frac{\theta + 2\pi k}{2}} +{\displaystyle \mathcal{O}(m)} & \text{(small $m$)}\\
\\
{\displaystyle -\frac{\chi_\text{pure} }{f_\pi^2 \chi m} } (\theta + 2\pi k) 
+ \mathcal{O}(m^{-2}) & \text{(large $m$)}, 
\end{cases}
\end{equation}
which give the potentials in these limits:
\begin{equation}\label{eq:Vphin}
\begin{split}
V(\phi , \phi + 2 \pi k) =
\begin{cases}
-2 f_\pi ^2 \chi m \cos \left(\displaystyle{ \frac{\theta + 2 \pi k}{2} }\right) + \mathcal{O}(m^2) \quad \quad \quad \quad \ \text{(small $m$)}
\\
\\
-2 f_\pi ^2 \chi m +{\displaystyle \frac{\chi_\mathrm{pure} }{2} }(\theta + 2\pi k )^2 
+\mathcal{O}\left( m^{-1}\right) \quad \quad \text{(large $m$)}.
\end{cases}
\end{split}
\end{equation}
\\

\begin{figure}
\begin{center}
\includegraphics[width=0.5\columnwidth]{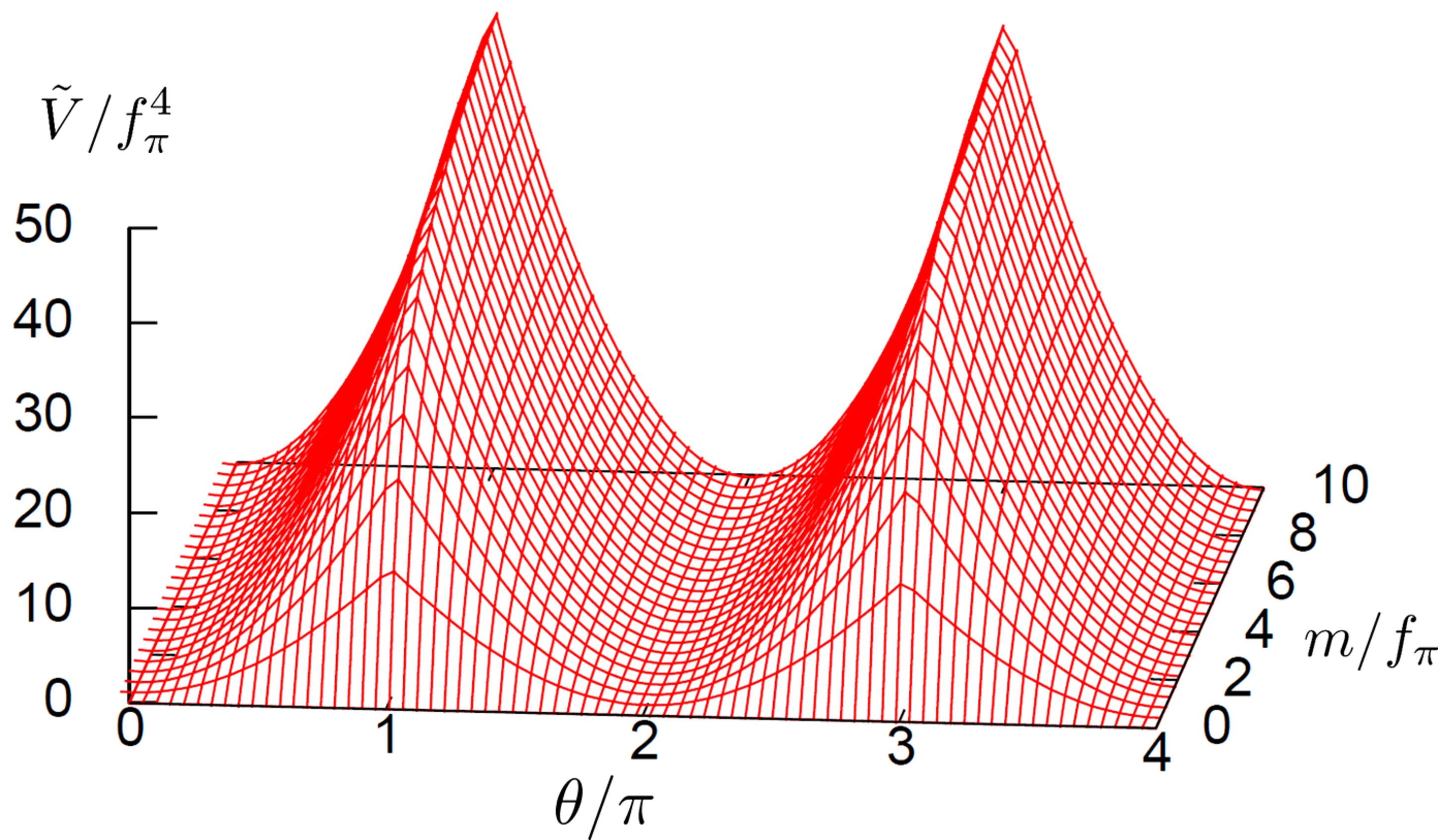}
 \caption{
QCD vacuum energy as a function of the $\theta$-angle and quark mass. The $\theta$-vacuum structure indicates that there are phase transitions at $\theta = (2n - 1)\pi $ unless quarks are massless. Here, we expressed $\tilde V = V + 2 f_\pi^2 \chi m $ instead of $V$ so that the $\theta$-dependence appears more clearly. Also we chose the parameters as $f_\pi = 93\ {\rm MeV}$, $f_\pi^2 \chi = -\langle\bar{q}q\rangle = (250\ \text{MeV})^3$, and $\chi_\text{pure}=(170\ \text{MeV})^4$.
}
 \label{fig:vac}
\end{center}
\end{figure}

Note that the potential $V(\phi, \phi + 2 \pi k )$ is a candidate of the true ground state energy. In other words, $V(\phi, \phi + 2 \pi k )$ is regarded as the potential of a metastable vacuum state labeled with an integer $k$, and the true vacuum energy is given by $ \min_k \left[ V(\phi, \phi+2\pi k) \right ] $. (In Sec.~\ref{sec:ms}, we will give a detailed discussion concerning these metastable states.) As shown in Ref.~\cite{Witten:1980sp}, the vacuum for small mass is decided by the competition between two metastable state with even $k$ and the one with odd $k$. On the other hand, we also obtain the same result as the one derived from the holographic approach~\cite{Witten:1998uka}; the vacuum of the pure Yang--Mills theory is determined by the minimum of the parabolic functions, i.e. $\min_k (\theta + 2 \pi k )^2$. As mentioned in Sec.~\ref{sec:VD}, the Di-Vecchia--Veneziano model is based on the chiral perturbation theory, and thus naively it might seem that this model cannot be directly applied to cases where quarks are not light. Nevertheless, we stress that our extension of this model is validated because the vacuum in the pure Yang--Mills theory is derived within this model. Additionally, if this model can be used even for arbitrary quark mass, we obtain the vacuum energies as a function of quark mass and the $\theta$-angle, from numerical calculation of Eq.~\eqref{eq:minimize} (see Fig.~\ref{fig:vac}). This figure shows that the Di-Vecchia--Veneziano model gives the $\theta$-vacua as a continuous function of quark mass, and that it reproduces the characteristic first order phase transition at $\theta = \pi \ ({\rm mod}\ 2 \pi)$, unless quarks are massless.

We understand why the vacuum energies for small and large mass are described as different functions of the $\theta$-angle, considering the structure of Eq.~\eqref{eq:eqk} and its solution Eq.~\eqref{eq:solution}. As mass increased, the solution $\phi (\theta, m)$ becomes much smaller, and eventually approaches to zero, as described in Eq.~\eqref{eq:solution}. In other words, quark mass plays the role of fixing the dynamical meson field $U$ to unity. In this sense, we find an answer for the question raised in Sec.~\ref{sec:mdth}: the reason why the pure Yang--Mills theory with the Chern--Simons term $iQ\theta$ cannot be obtained by integrating out the ${\rm U(1)_A}$-rotated heavy fermion field $e^{i \gamma_5 \theta / 2\Nf } q$. Namely, for heavy quarks, a change of the phase of $U$, such as an axial rotation, is prohibited energetically because such a transformed $U$ does not minimize the potential energy.

At the end of this section, we comment about the periodicity of $\theta$. Any physical quantity should be a function with the periodicity under the transformation $\theta\to \theta + 2\pi$. We have to note that the periodicity of $\theta$ is necessary not for the energy of a metastable state, but for the vacuum energy. In fact, its potentials Eq.~\eqref{eq:Vphin} are not such periodic function. The periodicity in terms of the vacuum energy is verified, if one notices from Eq.~\eqref{eq:eqk} that the transformation $\theta\to\theta + 2\pi$ is equivalent to the change of an integer $k$ into $k+1$. Therefore, the vacuum energy given by $\min_k [V(\phi,\phi+2\pi k)]$ does not change under the transformation $k\to k + 1$, or equivalently $\theta \to \theta + 2\pi$.

\section{Chiral condensate}\label{sec:cc}
\begin{center}
\begin{figure}
\begin{minipage}{0.48\hsize}
\begin{center}
\includegraphics[width=1\columnwidth]{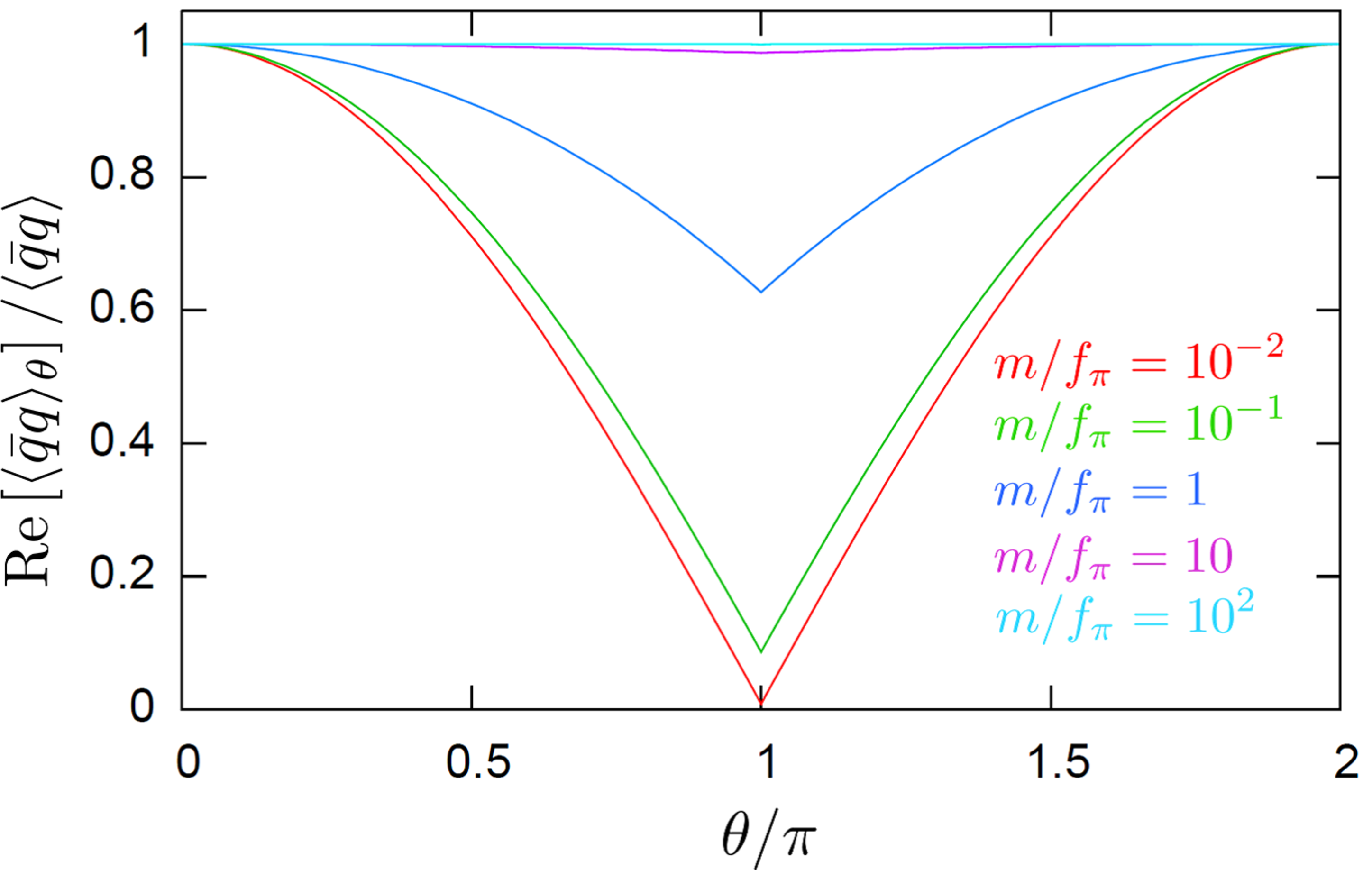}
\end{center}
\end{minipage}
\hspace{0.04\hsize}
\begin{minipage}{0.48\hsize}
\vspace{-5pt}
\begin{center}
\includegraphics[width=1\columnwidth]{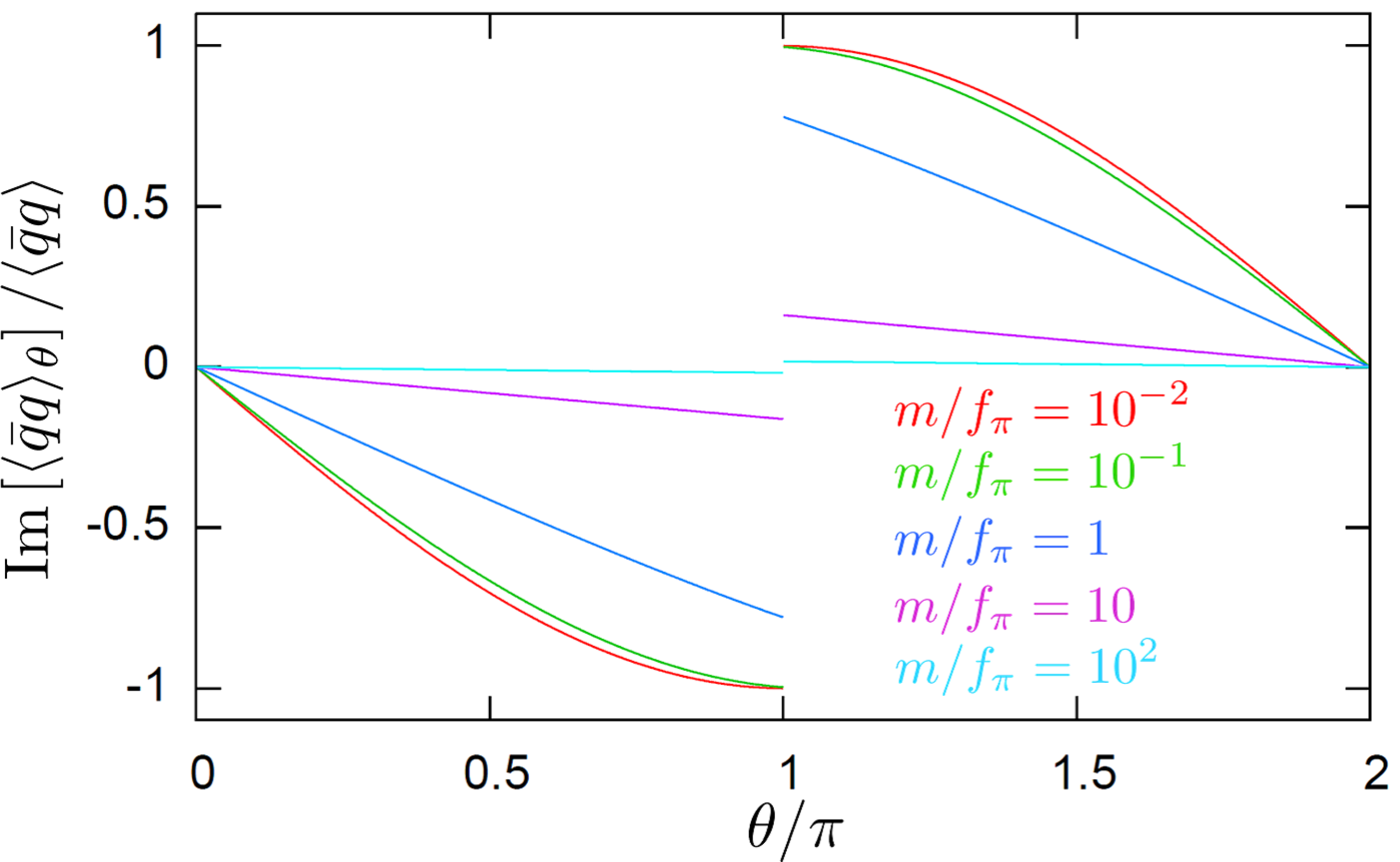}
 \end{center}
\end{minipage}
\caption{
$\theta$- and mass-dependence of the complex chiral condensate. For both the real and imaginary part of the condensate, there are singularities at $\theta = \pi$. The same result is pointed out for small mass in Ref.~\cite{Boer:2008ct}. These condensates for large mass do not depend on the $\theta$-angle, and especially the imaginary part is suppressed. This is because the phases of meson field $U$, i.e. $\phi_{u}$ and $\phi_{d}$ are fixed dynamically to zero for heavy quark mass. Therefore, the chiral condensate for large mass is fixed to the real value $\langle \bar q q \rangle$ even if $\theta \neq 0$.
}
\label{fig:ccc}
\end{figure}
\end{center}

As well known, the chiral condensate with a finite $\theta$-angle becomes a complex value, because under the ${\rm U(1)_A}$-rotation with the angle $\theta / 2\Nf$, the mass term is transformed as
\begin{equation}\label{eq:ccc}
\begin{split}
m \bar q q &\to 
m \bar q e^{i\gamma_5\theta / \Nf}q \\
&= m \bar q q \cos(\theta /\Nf) 
+i m \bar q\gamma_5 q  \sin(\theta /\Nf).
\end{split}
\end{equation}
Since in our argument, chiral condensate is regarded as ${\rm tr} \ U$ with phases $\phi_u$ and $\phi_d$, its real part (or $\sigma$-condensate) and imaginary part (or $\eta$-condensate) can be obtained directly from the solution $\phi (\theta, m)$ for the minimized equation Eq.~\eqref{eq:eqk};
\begin{equation}\label{eq:sigmaeta}
{\rm Re}\langle \bar qq \rangle_\theta = \langle \bar q q \rangle \cos \phi(\theta, m), \quad
{\rm Im}\langle \bar qq \rangle_\theta = \langle \bar q q \rangle \sin \phi(\theta, m) ,
\end{equation}
where $\langle \bar q q \rangle$ stands for the chiral condensate for $\theta=0$. The mass- and $\theta$-dependences of the condensates are shown in Fig.~\ref{fig:ccc}. From the chiral condensate as a function of quark mass and the $\theta$-angle, we confirm again the contribution of mass to the $\theta$-vacua, and also obtain a new suggestion in terms of the $\theta$-dependence of chiral condensate. As discussed in Sec.~\ref{sec:anly}, the solution $ \phi (\theta, m)$ for large mass is fixed to zero. As a result of this, the imaginary part of the complex chiral condensate $\sim \sin \phi (\theta, m)$ is suppressed for large quark mass (see the right panel of Fig.~\ref{fig:ccc}). This means that the complex chiral condensate, even for finite $\theta$-angle, becomes a real value in the large mass limit. In other words the complex chiral condensate is also fixed to the real direction on the chiral circle, due to the effect of large mass. Besides, based on such a fixation mechanism of the chiral condensate, we can obtain another reason why taking the quenched limit after the axial rotation Eq.~\eqref{eq:ccc} leads to an incorrect action without the topology. Namely, the real direction of chiral condensate on the complex plane is most favorable, and thus the axial rotation in the large mass limit is energetically forbidden.

For light quarks, we also find the typical behavior in Fig.~\ref{fig:ccc}. As long as quarks are light enough, the $\theta$-dependence of the imaginary part remains because there is no energetic restriction of the axial rotation. Moreover, the singularities in Fig.~\ref{fig:ccc} are related to that of the potential Fig.~\ref{fig:vac}. Indeed it is pointed out that the Numbu--Jona-Lasinio model leads to such a singularity of chiral condensate at $\theta = \pi $~\cite{Boer:2008ct} (see also Ref.~\cite{Hosotani:1996hy}, which gives a similar discussion of the massive Schwinger model).

Based on the periodicity of the $\theta$-angle for the vacuum energy, we can check that for the chiral condensate. As mentioned in Sec.~\ref{sec:anly}, the vacuum energy given by the true solution for Eq.~\eqref{eq:minimize} is a periodic function of the $\theta$-angle. This implies that the true solution should also be periodic under the transformation $\theta\to \theta + 2\pi$. On the other hand, the chiral condensate is also determined by such a solution, through Eq.~\eqref{eq:sigmaeta}. Thus, the chiral condensate is, actually, periodic under the transformation $\theta\to\theta + 2\pi$.

\section{Metastable states}\label{sec:ms}
In this section, we discuss metastable states that are the origin of the first order phase transitions at $\theta = \pi$. In order to do this, first we review a remarkable work related to the nature of the metastable state in QCD~\cite{Azcoiti:2003ai}.

The QCD partition function including the $\theta$-angle can be written in the following form,
\begin{equation}\label{eq:z}
Z(\theta) = \sum_Q Z_Q e^{iQ\theta} = \sum_{\tilde Q} f(\tilde Q) e^{i\theta V_4 \tilde Q}.
\end{equation}
Here, we introduced the topological charge density $\tilde Q = Q / V_4$, and $V_4$ denotes the four dimensional volume. (Note that we used and will use some different notations from those of the original paper.) This partition function, taking the large volume limit i.e., $\tilde Q \to x$, gives its continuous version;
\begin{equation}\label{eq:zc}
Z(\theta) \underset{V_4\to \infty}{\longrightarrow} {\cal Z}_{\rm c} (\theta) = V_4 \int dx f( x ) e^{i\theta V_4 x}.
\end{equation}
Then using the Poisson summation formula, one can obtain the relation between $Z(\theta)$ and ${\cal Z}_{\rm c}(\theta)$ as
\begin{equation}\label{eq:zcsum}
Z(\theta) = \sum_n {\cal Z}_{\rm c} (\theta + 2\pi n) .
\end{equation}
This relation implies a physical meaning of the integers $n$. In the large mass limit, the sum in Eq.~\eqref{eq:zcsum} may receive only one dominant contribution, and thus the QCD ground potential takes the following form; 
\begin{equation}
\Omega = - \frac{1}{V_4}\ln Z(\theta) \underset{V_4\to \infty}{\longrightarrow} 
\min_{n}\left[ -\frac{1}{V_4} \ln {\cal Z}_{\rm c} (\theta + 2 \pi n) \right ] .
\end{equation}
In this sense, ${\cal Z}_{\rm c}(\theta + 2 \pi n)$ is regarded as the partition function of the metastable state labeled by $n$. Note that an integer $n$ has nothing to do with the label related to topological vacua, i.e. $Q$ or $x$. Indeed the partition function of $n$-state includes the contributions from all topological vacua, as shown in Eq.~\eqref{eq:zc}.

\begin{center}
\begin{figure}[t]
\begin{center}
\includegraphics[width=0.5\columnwidth]{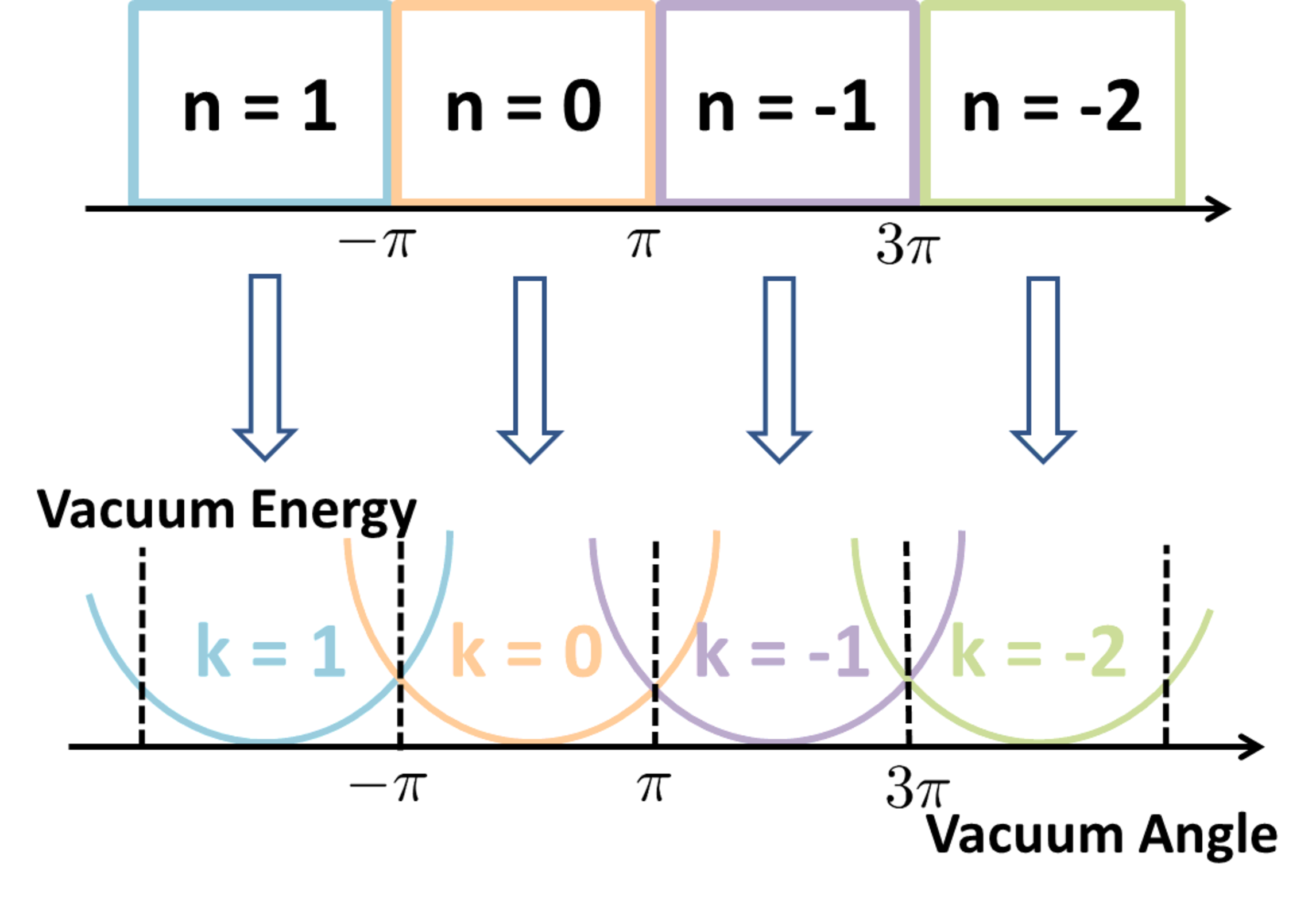} 
 \caption{
Interpretation of the phase transition in the pure Yang--Mills theory, based on Ref.~\cite{Azcoiti:2003ai}. In this case, all the domains in the upper figure correspond to each parabola, and the metastable states are labeled by integer $k$ in Eq.~\eqref{eq:eqk}. The phase transition $\theta = \pi $ for heavy quarks results from the infinite parabolic energy functions.
}
 \label{fig:degePYM}
\end{center}
\end{figure}
\end{center}

\begin{center}
\begin{figure}[t]
\begin{center}
 \includegraphics[width=0.5\columnwidth]{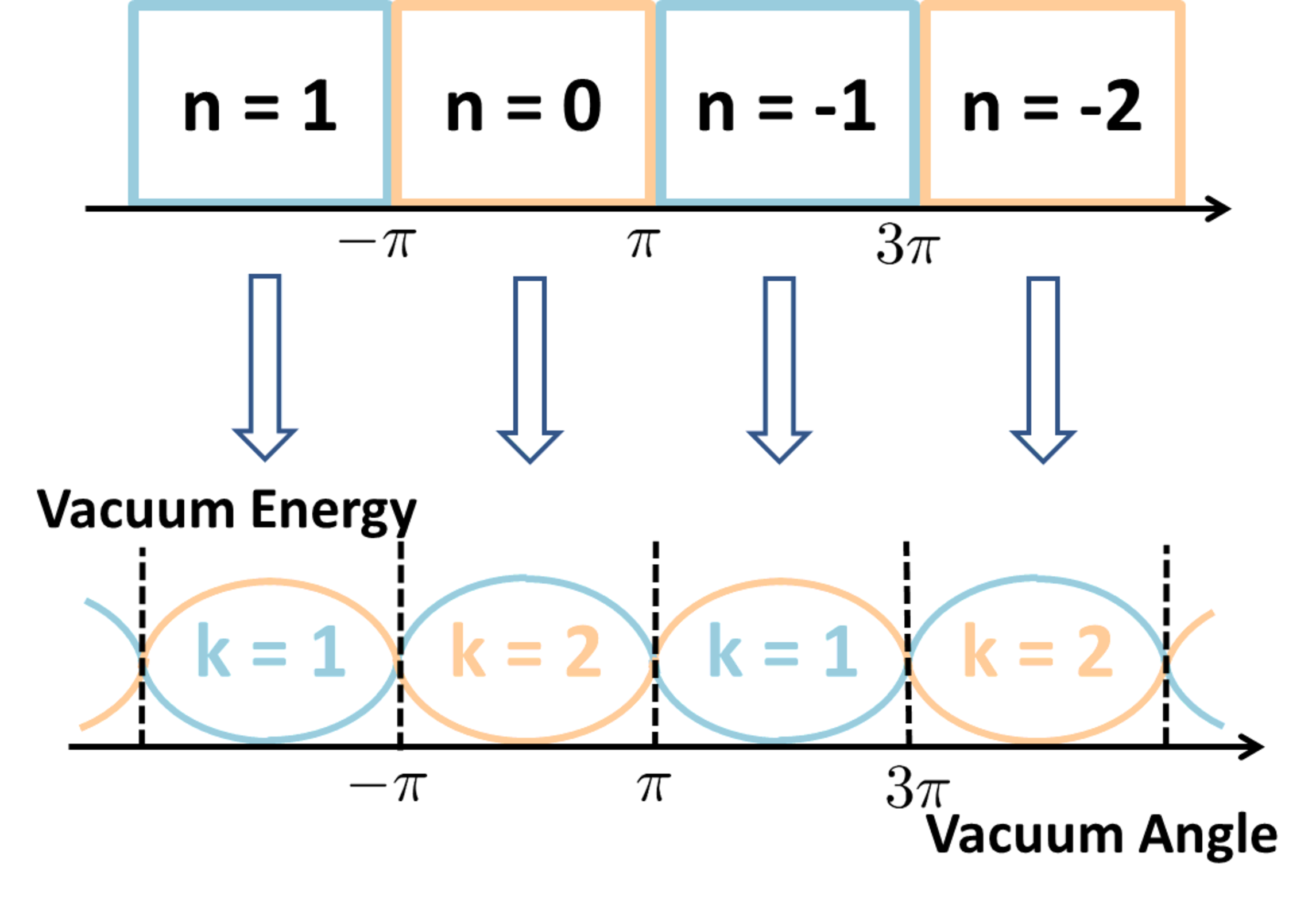}
 \caption{
Degeneracy of the metastable states for light quarks. The true vacuum energy Eq.~\eqref{eq:chie} for $\Nf = 2$ consists of only two energy functions: $ -\cos \left ( \theta /2 + \pi \right )$ for $k = 1$ and $ -\cos \left ( \theta / 2 \right )$ for $k = 2$. All the metastable states labeled by odd (even) $n$ degenerate as the $k = 1$ ($k = 2$) states.
}
 \label{fig:degechi}
 \end{center}
\end{figure}
\end{center}

In this context, let us consider \calCP symmetry. Similarly to the partition function Eq.~\eqref{eq:zcsum}, the topological charge density, which is the order parameter of \calCP symmetry, is also given by the sum of the contribution from each metastable state;
\begin{equation}\label{eq:xsum}
\begin{split}
\langle \tilde Q \rangle = \sum_n \langle x_n \rangle, 
\quad \langle x_n \rangle = \frac{V_4}{Z(\theta)} \int dx x f(x) e^{i(\theta + 2\pi n) x V_4}.
\end{split} 
\end{equation}
The topological charge density $\langle \tilde Q \rangle $ should vanish at $\theta = \pi$ because there is the \calCP symmetry for $\theta = \pi$. In fact, although each $\langle x_n \rangle_{\theta = \pi}$ my be a finite value, all adjacent pairs cancel out; based on the definition of $\langle x_n \rangle $ in Eq.~\eqref{eq:xsum}, we can check $\langle x_n \rangle_{\theta = \pi} = -\langle x_{n-1} \rangle_{\theta = \pi}$. In the large volume limit, however, \calCP might be broken. This is because for large volume, only one dominant metastable state remains in the sum Eq.~\eqref{eq:zcsum} and so in Eq.~\eqref{eq:xsum}. It is possible, hence, that the total topological charge density $\langle \tilde Q \rangle$ becomes finite, as long as that in the dominant state is not zero. We note that a dominant metastable state is selected by the value of the $\theta$-angle. Suppose that the $n=0$ metastable state is dominant for $0 \leq \theta \leq \pi$ (at least, it is dominant for small $\theta$ since all the others are suppressed by the phase factor $e^{i 2\pi n x V_4}$). In this case, the $n=-1$ state also becomes dominant for $\pi \leq \theta \leq 3\pi$. Thus, $\theta = \pi$ could be the boundary between the phases dominated by the $n=0$- and $n=-1$ state. This is the main idea of origin of the phase transition at $\theta = \pi$. In the upper parts of Fig.~\ref{fig:degePYM} and Fig.~\ref{fig:degechi}, these domains are shown.

For the investigation of the metastable states in the Di-Vecchia--Veneziano model, we mention the following three remarks (the first and second are actually connected to our discussion). 
First, each domain does not have to be physically different from each other, and even if several phases are equivalent, the phase transition might occur at $\theta = \pi$. For example, suppose that all the phases dominated by an odd $n$-state are equivalent and so are all the phases dominated by an even $n$-state, and that these two types of domains are different. Namely, there are two independent phases. Then the phase transition at $\theta =\pi$ exists since $\theta = \pi$ is the boundary between different phases dominated by the $n=0$- and $n=-1$ state. Secondly, even if one takes the large volume limit, the phase transition at $\theta = \pi$ does not always occur. There is no phase transition at $\theta = \pi$, for example, if the $n = 0$ and $n = -1$ state are physically equivalent. Hence, there is the first order phase transition at $\theta = \pi$ only if $\langle x_0 \rangle _{\theta = \pi} \neq \langle x_{-1} \rangle _{\theta = \pi}$. Finally, we also have to note that the aforementioned mechanism of the phase transition is rather general, except for an assumption that the $n=0$ state is dominant for $0 \leq \theta \leq \pi$. In other words, even for the Di-Vecchia--Veneziano model, the origin of the first order phase transition at $\theta = \pi$ should be explained, considering the innate metastable states included in QCD.
\\

Next we analyze the relation between quark mass and the metastable states. In the pure gluonic theory, the energy density of the $\theta$-vacuum is given by~\cite{Witten:1998uka},
\begin{equation}\label{eq:yme}
E(\theta) \propto \min _k (\theta + 2\pi k)^2.
\end{equation}
From the viewpoint of the metastable states, this energy configuration means that the integer $k$ in Eq.~\eqref{eq:yme} corresponds one-to-one with the label of these states, $n$ in Eq.\eqref{eq:zcsum} (see Fig.~\ref{fig:degePYM}). We can confirm directly, thus, that the phase transition for large mass is caused by the metastable states. On the other hand, for small mass, there are only a few types of the metastable states~\cite{Witten:1980sp}. Indeed Eq.~\eqref{eq:Vphin} shows that the potential energies for small mass are separated by the two types corresponding to odd $k$ and even $k$. More generally, the true vacuum for the $\Nf$ light flavors case consists of $\Nf$ types of potential, and the vacuum energy is given by
\begin{equation}\label{eq:chie}
E(\theta) \propto \min_{k = 1, \cdots, \Nf} \left [-\cos \left( \frac{\theta +2 \pi k}{\Nf} \right ) \right ].
\end{equation}
Since $k$ never coincide with $n$ in Eq.~\eqref{eq:zcsum}, it seems naively that the phase transition for light mass should be interpreted as physically different from that in the pure Yang--Mills theory. The phase transition can also be explained by the metastable states, however, if we consider that an infinite number of metastable states are degenerate (see Fig.~\ref{fig:degechi}). Namely, the phase transition for small mass results from the degeneracy of the metastable states; $\big \{ n = 0, \pm 1, \pm 2, \cdots \big \}\longrightarrow \big \{ k = 1, 2, \cdots, \Nf \}$.


\begin{figure}[t]
 \includegraphics[width=1\columnwidth]{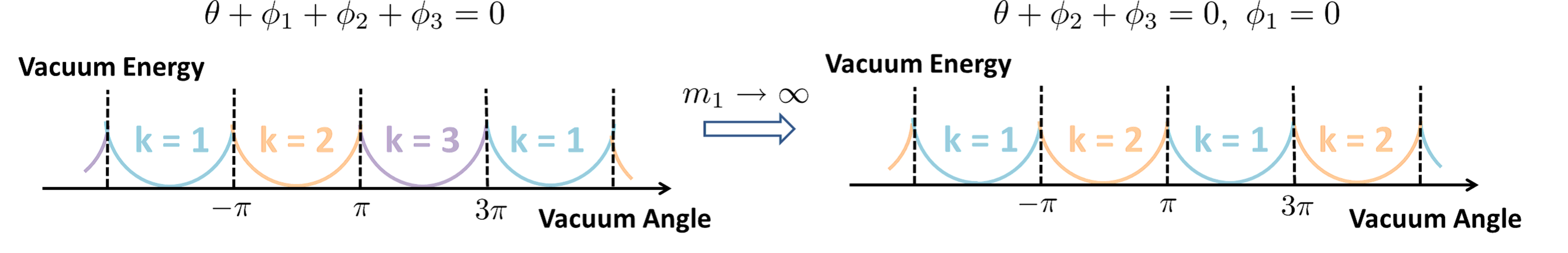}
 \caption{Metastable states under the presence of three light flavors (Left), and one heavy and two light flavors (Right). Because heavy flavor is fixed to zero, as mentioned in Sec.~\ref{sec:anly}, only the light flavors contribute to the equation of motion. As a result of this, the vacuum energy of the case described in the Right panel is equivalent to the one for $\Nf = 2$. One can confirm from the example, therefore, that the degeneracy factor of the metastable states is determined not by the number of flavors included in the theory, but the one of the dynamical flavors.}
 \label{fig:1hv2li}
\end{figure}

Here, analyzing again the equation of motion of the Di-Vecchia--Veneziano model, we can clarify the structure of the degenerate metastable states. If we suppose that all flavors are light, the solution is given approximately by the simple equation of motion
\begin{equation}\label{eq:topeq}
\theta + \sum_i \phi_i =0,
\end{equation}
so that the topological term $\sim ( \theta + \sum_i \phi_i )^2$ in the potential should be very small. Considering heavy quarks together with light ones, however, Eq.~\eqref{eq:topeq} should be changed. For example, let us assume that there are one heavy flavor named $\phi_1$, and two light flavors, $\phi_2$ and $\phi_3$. As we see in Sec.~\ref{sec:anly}, the heavy flavor $\phi_1$ is fixed to zero so that the mass term becomes small. The equation of motion to solve, therefore, is rewritten as
\begin{equation}
\theta + \phi_2 + \phi_3 = 0, \quad \quad \phi_1 = 0.
\end{equation}
As a result, the $\theta$-dependence is determined only by the two light flavors. As shown in Fig.~\ref{fig:1hv2li}, this is equivalent to the case that there are only two dynamical flavors. The degeneracy factor of the metastable states is equal not to the number of the flavors included originally in theory, but to that of dynamical quarks.

\begin{figure}[t]
\begin{center}
 \includegraphics[width=0.5\columnwidth]{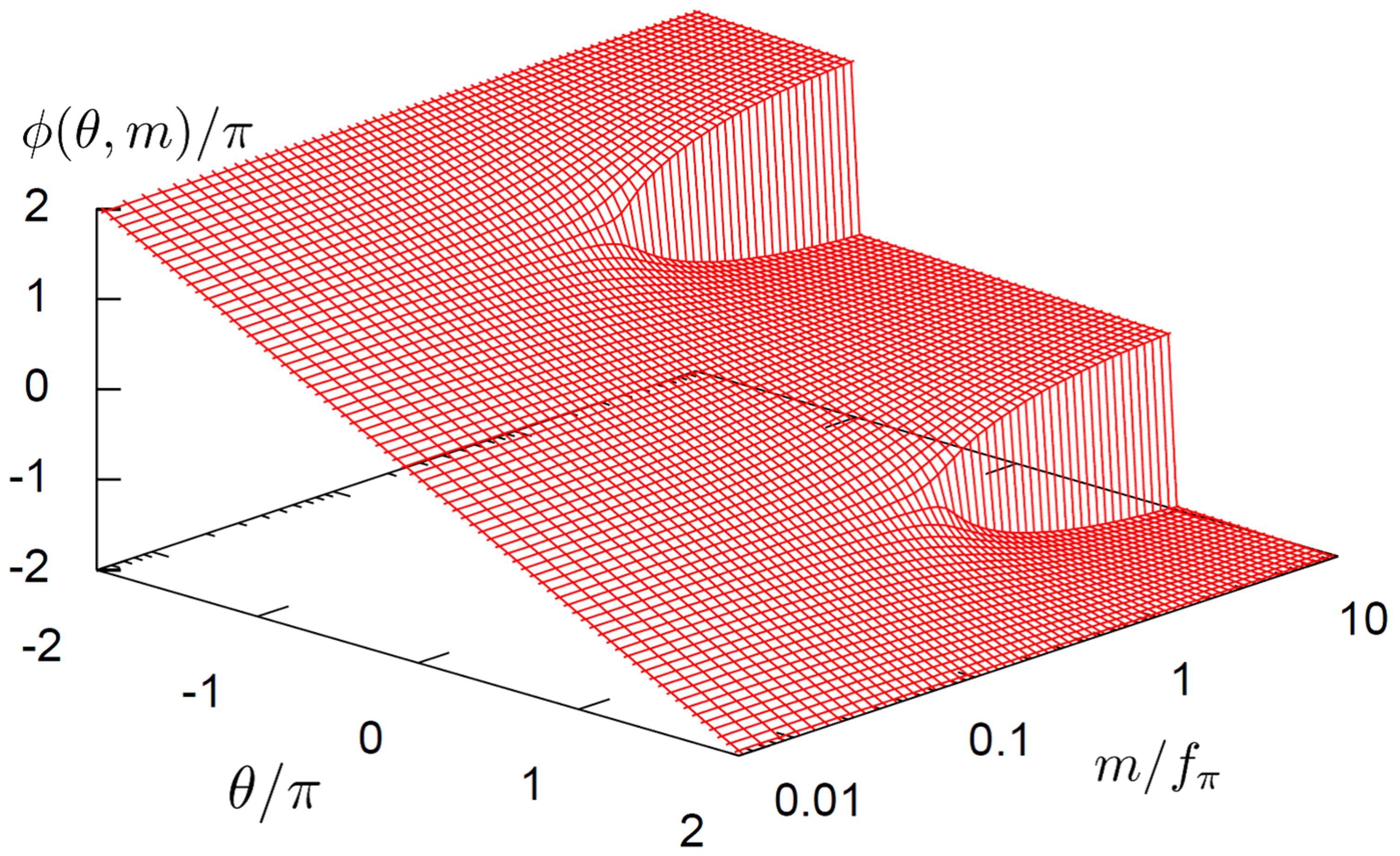}
 \caption{Classical solution of the one flavor Di-Vecchia--Veneziano model as a function of the $\theta$-angle and quark mass. For small mass, the solution is a smooth function of $\theta$, and thus there is no phase transition related to the variation of the $\theta$-angle. As mass increased, the discontinuity at $\theta = \pi$, however, appears in the vicinity of $m \simeq f_\pi$. This shows the existence of the crossover between the phases in which the \calCP at $\theta = \pi$ is restored (for small mass) and broken spontaneously (for large mass).
}
 \label{fig:solnf1}
\end{center}
\end{figure}

The above discussion leads us to an interesting result for the one flavor case. From the viewpoint of the degeneracy of the metastable states, there is no phase transition for the one light flavor case because all adjacent regions are dominated by the same metastable states. On the other hand, since the potential in the large mass limit is given by Eq.~\eqref{eq:yme} independently on $\Nf$, there appears the phase transition at $\theta = \pi$. For $\Nf=1$, this implies the presence of a new transition (strictly speaking, a crossover) between two phases in terms of the absence and the existence of the spontaneous \calCP breaking at $\theta = \pi$. If the solution $\phi_1 = \phi(\theta,m)$ has the singularity at $\theta = \pi$, the spontaneous \calCP breaking occurs at $\theta = \pi$. In Fig.~\ref{fig:solnf1}, therefore, we can confirm such a new phase transition, from the discontinuity of $\phi_1 = \phi(\theta,m)$ at $m \gtrsim f_\pi$. The same result is argued in Ref.~\cite{Creutz:1994px,Creutz:1995wf}.

\section{Conclusion}
We showed in this paper how quark mass contributes to the QCD $\theta$-vacua as a function of the $\theta$-angle and quark mass, within the Di-Vecchia--Veneziano model. The discussions in this paper are based of a new interpretation that the physics even for large mass is described by the Di-Vecchia--Veneziano model. This is confirmed, for example, from the fact that the topological susceptibility computed within this model reproduces an expected result for large mass; $\chi_{\rm top}(m) \to \chi_{\rm pure}$. First of all, we investigated the effect of heavy quark to the $\theta$-vacuum state, calculating the potential energy of the vacuum. Then we found that the phase of the meson field minimizing the potential is fixed to zero. As a result of this, any axial rotation for large mass is forbidden energetically. This argument leads to the mass dependence of chiral condensate; as mass increased, the chiral condensate after the axial rotation, i.e., $\langle \bar q e^{i \gamma_5 \theta / \Nf} q \rangle$ is fixed to the real direction on the chiral circle.

We analyzed the innate metastable states included in the QCD $\theta$-vacuum (these are different from the topological vacua labeled by the Chern--Simons number). In the Veneziano--Di- Vecchia model, there are basically infinite solutions for the classical equation of motion, and each solution is labeled by an integer $k$. In the quenched limit, all solutions give different potential energies, and thus the vacuum is decided by the infinite metastable states. In contrast, there are only $\Nf$ solutions in the small mass limit. As a result, the vacuum state for small mass consists of $\Nf$ metastable states. The reduction of the number of the metastable states as quark mass decreased implies that dynamical quarks play an important role leading to the degeneracy of the metastable states underlying full QCD. Indeed, we checked that for $\Nf=1$, a unique metastable state in the small mass limit is split into a number of metastable states, as mass increased. Additionally, the phase transitions at $\theta = \pi$ for small and large mass are also interpreted uniformly; for arbitrary mass, $\theta = \pi$ is the boundary between the two phases dominated by different metastable states.

Finally, we stress that a thorough investigation of the property of the $\theta$-vacua, especially in regard to the mass-dependence. For example, we need to discuss the heavy quark effective theory with the $\theta$-vacuum taken account into. Such an approach might give supports for our calculation based on the Di-Vecchia--Veneziano model. At the same time, the numerical method instead of lattice QCD simulation, which does not work for finite $\theta$-angle, also has to be performed for the study of the $\theta$-vacuum structure~\cite{Aarts:2010gr,Aarts:2013nja}. Besides, combining this new effective theory with the locally \calP and/or \calCP violating domains in the collision~\cite{Morley:1983wr,Kharzeev:1998kz}, the discussion of the $\theta$-dependence for large mass would be quite fruitful for the heavy flavors, which are the important probes in order to identify the initial state of the relativistic heavy ion collision. Conversely, from a new actual estimation of the energy loss of the heavy quark jet~\cite{Braaten:1991we,Herzog:2006gh} with consideration of the $\theta$-vacua for large mass, one could obtain some information of the locally \calP and/or \calCP violating bubbles in the collision.

\vspace{10pt}
The author is grateful to Kenji Fukushima, Arata Yamamoto, Pablo A. Morales, and Sanjin Benic for valuable discussions. The author also thanks the Yukawa Institute for Theoretical Physics, Kyoto University, where this work was developed considerably during the YITP-T-13-05 on ``New Frontiers in QCD 2013".



\begin{thebibliography}{10}
\expandafter\ifx\csname url\endcsname\relax
  \def\url#1{\texttt{#1}}\fi
\expandafter\ifx\csname urlprefix\endcsname\relax\def\urlprefix{URL }\fi
\expandafter\ifx\csname href\endcsname\relax
  \def\href#1#2{#2} \def\path#1{#1}\fi

\bibitem{Baker:2006ts}
C.~Baker, D.~Doyle, P.~Geltenbort, K.~Green, M.~van~der Grinten, et~al., {An
  Improved experimental limit on the electric dipole moment of the neutron},
  Phys.Rev.Lett. 97 (2006) 131801.
\newblock \href {http://arxiv.org/abs/hep-ex/0602020}
  {\path{arXiv:hep-ex/0602020}}, \href
  {http://dx.doi.org/10.1103/PhysRevLett.97.131801}
  {\path{doi:10.1103/PhysRevLett.97.131801}}.

\bibitem{Cheng:1987gp}
H.-Y. Cheng, {The Strong CP Problem Revisited}, Phys.Rept. 158 (1988) 1.
\newblock \href {http://dx.doi.org/10.1016/0370-1573(88)90135-4}
  {\path{doi:10.1016/0370-1573(88)90135-4}}.

\bibitem{Peccei:1998jt}
R.~Peccei, {Reflections on the strong CP problem}\href
  {http://arxiv.org/abs/hep-ph/9807514} {\path{arXiv:hep-ph/9807514}}.

\bibitem{Witten:1979vv}
E.~Witten, {Current Algebra Theorems for the U(1) Goldstone Boson}, Nucl.Phys.
  B156 (1979) 269.
\newblock \href {http://dx.doi.org/10.1016/0550-3213(79)90031-2}
  {\path{doi:10.1016/0550-3213(79)90031-2}}.

\bibitem{Veneziano:1979ec}
G.~Veneziano, {U(1) Without Instantons}, Nucl.Phys. B159 (1979) 213--224.
\newblock \href {http://dx.doi.org/10.1016/0550-3213(79)90332-8}
  {\path{doi:10.1016/0550-3213(79)90332-8}}.

\bibitem{Urban:2009vy}
F.~R. Urban, A.~R. Zhitnitsky, {The cosmological constant from the QCD
  Veneziano ghost}, Phys.Lett. B688 (2010) 9--12.
\newblock \href {http://arxiv.org/abs/0906.2162} {\path{arXiv:0906.2162}},
  \href {http://dx.doi.org/10.1016/j.physletb.2010.03.080}
  {\path{doi:10.1016/j.physletb.2010.03.080}}.

\bibitem{Urban:2009yg}
F.~R. Urban, A.~R. Zhitnitsky, {The QCD nature of Dark Energy}, Nucl.Phys. B835
  (2010) 135--173.
\newblock \href {http://arxiv.org/abs/0909.2684} {\path{arXiv:0909.2684}},
  \href {http://dx.doi.org/10.1016/j.nuclphysb.2010.04.001}
  {\path{doi:10.1016/j.nuclphysb.2010.04.001}}.

\bibitem{Ohta:2010in}
N.~Ohta, {Dark Energy and QCD Ghost}, Phys.Lett. B695 (2011) 41--44.
\newblock \href {http://arxiv.org/abs/1010.1339} {\path{arXiv:1010.1339}},
  \href {http://dx.doi.org/10.1016/j.physletb.2010.11.044}
  {\path{doi:10.1016/j.physletb.2010.11.044}}.

\bibitem{Morley:1983wr}
P.~Morley, I.~Schmidt, {Strong P, {CP}, T Violations in Heavy Ion Collisions},
  Z.Phys. C26 (1985) 627.
\newblock \href {http://dx.doi.org/10.1007/BF01551807}
  {\path{doi:10.1007/BF01551807}}.

\bibitem{Kharzeev:1998kz}
D.~Kharzeev, R.~Pisarski, M.~H. Tytgat, {Possibility of spontaneous parity
  violation in hot QCD}, Phys.Rev.Lett. 81 (1998) 512--515.
\newblock \href {http://arxiv.org/abs/hep-ph/9804221}
  {\path{arXiv:hep-ph/9804221}}, \href
  {http://dx.doi.org/10.1103/PhysRevLett.81.512}
  {\path{doi:10.1103/PhysRevLett.81.512}}.

\bibitem{Peccei:1977hh}
R.~Peccei, H.~R. Quinn, {CP Conservation in the Presence of Instantons},
  Phys.Rev.Lett. 38 (1977) 1440--1443.
\newblock \href {http://dx.doi.org/10.1103/PhysRevLett.38.1440}
  {\path{doi:10.1103/PhysRevLett.38.1440}}.

\bibitem{Vilenkin:1980fu}
A.~Vilenkin, {EQUILIBRIUM PARITY VIOLATING CURRENT IN A MAGNETIC FIELD},
  Phys.Rev. D22 (1980) 3080--3084.
\newblock \href {http://dx.doi.org/10.1103/PhysRevD.22.3080}
  {\path{doi:10.1103/PhysRevD.22.3080}}.

\bibitem{Son:2004tq}
D.~Son, A.~R. Zhitnitsky, {Quantum anomalies in dense matter}, Phys.Rev. D70
  (2004) 074018.
\newblock \href {http://arxiv.org/abs/hep-ph/0405216}
  {\path{arXiv:hep-ph/0405216}}, \href
  {http://dx.doi.org/10.1103/PhysRevD.70.074018}
  {\path{doi:10.1103/PhysRevD.70.074018}}.

\bibitem{Metlitski:2005pr}
M.~A. Metlitski, A.~R. Zhitnitsky, {Anomalous axion interactions and
  topological currents in dense matter}, Phys.Rev. D72 (2005) 045011.
\newblock \href {http://arxiv.org/abs/hep-ph/0505072}
  {\path{arXiv:hep-ph/0505072}}, \href
  {http://dx.doi.org/10.1103/PhysRevD.72.045011}
  {\path{doi:10.1103/PhysRevD.72.045011}}.

\bibitem{Kharzeev:2007jp}
D.~E. Kharzeev, L.~D. McLerran, H.~J. Warringa, {The Effects of topological
  charge change in heavy ion collisions: 'Event by event P and CP violation'},
  Nucl.Phys. A803 (2008) 227--253.
\newblock \href {http://arxiv.org/abs/0711.0950} {\path{arXiv:0711.0950}},
  \href {http://dx.doi.org/10.1016/j.nuclphysa.2008.02.298}
  {\path{doi:10.1016/j.nuclphysa.2008.02.298}}.

\bibitem{Fukushima:2008xe}
K.~Fukushima, D.~E. Kharzeev, H.~J. Warringa, {The Chiral Magnetic Effect},
  Phys.Rev. D78 (2008) 074033.
\newblock \href {http://arxiv.org/abs/0808.3382} {\path{arXiv:0808.3382}},
  \href {http://dx.doi.org/10.1103/PhysRevD.78.074033}
  {\path{doi:10.1103/PhysRevD.78.074033}}.

\bibitem{Kharzeev:2010gd}
D.~E. Kharzeev, H.-U. Yee, {Chiral Magnetic Wave}, Phys.Rev. D83 (2011) 085007.
\newblock \href {http://arxiv.org/abs/1012.6026} {\path{arXiv:1012.6026}},
  \href {http://dx.doi.org/10.1103/PhysRevD.83.085007}
  {\path{doi:10.1103/PhysRevD.83.085007}}.

\bibitem{Dashen:1970et}
R.~F. Dashen, {Some features of chiral symmetry breaking}, Phys.Rev. D3 (1971)
  1879--1889.
\newblock \href {http://dx.doi.org/10.1103/PhysRevD.3.1879}
  {\path{doi:10.1103/PhysRevD.3.1879}}.

\bibitem{Coleman:1976uz}
S.~R. Coleman, {More About the Massive Schwinger Model}, Annals Phys. 101
  (1976) 239.
\newblock \href {http://dx.doi.org/10.1016/0003-4916(76)90280-3}
  {\path{doi:10.1016/0003-4916(76)90280-3}}.

\bibitem{Witten:1980sp}
E.~Witten, {Large N Chiral Dynamics}, Annals Phys. 128 (1980) 363.
\newblock \href {http://dx.doi.org/10.1016/0003-4916(80)90325-5}
  {\path{doi:10.1016/0003-4916(80)90325-5}}.

\bibitem{'tHooft:1981ht}
G.~'t~Hooft, {Topology of the Gauge Condition and New Confinement Phases in
  Nonabelian Gauge Theories}, Nucl.Phys. B190 (1981) 455.
\newblock \href {http://dx.doi.org/10.1016/0550-3213(81)90442-9}
  {\path{doi:10.1016/0550-3213(81)90442-9}}.

\bibitem{Ohta:1981ai}
N.~Ohta, {Vacuum Structure and Chiral Charge Quantization in the Large $N$
  Limit}, Prog.Theor.Phys. 66 (1981) 1408.
\newblock \href {http://dx.doi.org/10.1143/PTP.66.1408}
  {\path{doi:10.1143/PTP.66.1408}}.

\bibitem{Wiese:1988qz}
U.~Wiese, {Numerical Simulation of Lattice $\theta$ Vacua: The 2-$d$ U(1) Gauge
  Theory as a Test Case}, Nucl.Phys. B318 (1989) 153.
\newblock \href {http://dx.doi.org/10.1016/0550-3213(89)90051-5}
  {\path{doi:10.1016/0550-3213(89)90051-5}}.

\bibitem{Affleck:1991tj}
I.~Affleck, {Nonlinear sigma model at Theta = pi: Euclidean lattice formulation
  and solid-on-solid models}, Phys.Rev.Lett. 66 (1991) 2429--2432.
\newblock \href {http://dx.doi.org/10.1103/PhysRevLett.66.2429}
  {\path{doi:10.1103/PhysRevLett.66.2429}}.

\bibitem{Creutz:1994px}
M.~Creutz, {Chiral symmetry on the lattice}, Nucl.Phys.Proc.Suppl. 42 (1995)
  56--66.
\newblock \href {http://arxiv.org/abs/hep-lat/9411033}
  {\path{arXiv:hep-lat/9411033}}, \href
  {http://dx.doi.org/10.1016/0920-5632(95)00187-E}
  {\path{doi:10.1016/0920-5632(95)00187-E}}.

\bibitem{Creutz:1995wf}
M.~Creutz, {Quark masses and chiral symmetry}, Phys.Rev. D52 (1995) 2951--2959.
\newblock \href {http://arxiv.org/abs/hep-th/9505112}
  {\path{arXiv:hep-th/9505112}}, \href
  {http://dx.doi.org/10.1103/PhysRevD.52.2951}
  {\path{doi:10.1103/PhysRevD.52.2951}}.

\bibitem{Creutz:2009kx}
M.~Creutz, {Anomalies and chiral symmetry in QCD}, Annals Phys. 324 (2009)
  1573--1584.
\newblock \href {http://arxiv.org/abs/0901.0150} {\path{arXiv:0901.0150}},
  \href {http://dx.doi.org/10.1016/j.aop.2009.01.005}
  {\path{doi:10.1016/j.aop.2009.01.005}}.

\bibitem{Creutz:2010ee}
M.~Creutz, {Quark masses and strong CP violation}, AIP Conf.Proc. 1343 (2011)
  618--618.
\newblock \href {http://arxiv.org/abs/1011.0908} {\path{arXiv:1011.0908}},
  \href {http://dx.doi.org/10.1063/1.3575115} {\path{doi:10.1063/1.3575115}}.

\bibitem{Hosotani:1996hy}
Y.~Hosotani, R.~Rodriguez, J.~Hetrick, S.~Iso, {Confinement and chiral dynamics
  in the multiflavor Schwinger model}\href
  {http://arxiv.org/abs/hep-th/9606129} {\path{arXiv:hep-th/9606129}}.

\bibitem{Smilga:1998dh}
A.~V. Smilga, {QCD at theta similar to pi}, Phys.Rev. D59 (1999) 114021.
\newblock \href {http://arxiv.org/abs/hep-ph/9805214}
  {\path{arXiv:hep-ph/9805214}}, \href
  {http://dx.doi.org/10.1103/PhysRevD.59.114021}
  {\path{doi:10.1103/PhysRevD.59.114021}}.

\bibitem{Witten:1998uka}
E.~Witten, {Theta dependence in the large N limit of four-dimensional gauge
  theories}, Phys.Rev.Lett. 81 (1998) 2862--2865.
\newblock \href {http://arxiv.org/abs/hep-th/9807109}
  {\path{arXiv:hep-th/9807109}}, \href
  {http://dx.doi.org/10.1103/PhysRevLett.81.2862}
  {\path{doi:10.1103/PhysRevLett.81.2862}}.

\bibitem{Halperin:1998rc}
I.~E. Halperin, A.~Zhitnitsky, {Anomalous effective Lagrangian and theta
  dependence in QCD at finite N(c)}, Phys.Rev.Lett. 81 (1998) 4071--4074.
\newblock \href {http://arxiv.org/abs/hep-ph/9803301}
  {\path{arXiv:hep-ph/9803301}}, \href
  {http://dx.doi.org/10.1103/PhysRevLett.81.4071}
  {\path{doi:10.1103/PhysRevLett.81.4071}}.

\bibitem{Boer:2008ct}
D.~Boer, J.~K. Boomsma, {Spontaneous CP-violation in the strong interaction at
  theta = pi}, Phys.Rev. D78 (2008) 054027.
\newblock \href {http://arxiv.org/abs/0806.1669} {\path{arXiv:0806.1669}},
  \href {http://dx.doi.org/10.1103/PhysRevD.78.054027}
  {\path{doi:10.1103/PhysRevD.78.054027}}.

\bibitem{Boomsma:2009eh}
J.~K. Boomsma, D.~Boer, {The High temperature CP-restoring phase transition at
  theta = pi}, Phys.Rev. D80 (2009) 034019.
\newblock \href {http://arxiv.org/abs/0905.4660} {\path{arXiv:0905.4660}},
  \href {http://dx.doi.org/10.1103/PhysRevD.80.034019}
  {\path{doi:10.1103/PhysRevD.80.034019}}.

\bibitem{D'Elia:2012vv}
M.~D'Elia, F.~Negro, {$\theta$ dependence of the deconfinement temperature in
  Yang-Mills theories}, Phys.Rev.Lett. 109 (2012) 072001.
\newblock \href {http://arxiv.org/abs/1205.0538} {\path{arXiv:1205.0538}},
  \href {http://dx.doi.org/10.1103/PhysRevLett.109.072001}
  {\path{doi:10.1103/PhysRevLett.109.072001}}.

\bibitem{D'Elia:2013eua}
M.~D'Elia, F.~Negro, {Phase diagram of Yang-Mills theories in the presence of a
  $\theta$ term}, Phys.Rev. D88~(3) (2013) 034503.
\newblock \href {http://arxiv.org/abs/1306.2919} {\path{arXiv:1306.2919}},
  \href {http://dx.doi.org/10.1103/PhysRevD.88.034503}
  {\path{doi:10.1103/PhysRevD.88.034503}}.

\bibitem{Aoki:2014moa}
S.~Aoki, M.~Creutz, {Pion masses in 2-flavor QCD with $\eta$ condensation},
  Phys.Rev.Lett. 112 (2014) 141603.
\newblock \href {http://arxiv.org/abs/1402.1837} {\path{arXiv:1402.1837}},
  \href {http://dx.doi.org/10.1103/PhysRevLett.112.141603}
  {\path{doi:10.1103/PhysRevLett.112.141603}}.

\bibitem{Vicari:2008jw}
E.~Vicari, H.~Panagopoulos, {Theta dependence of SU(N) gauge theories in the
  presence of a topological term}, Phys.Rept. 470 (2009) 93--150.
\newblock \href {http://arxiv.org/abs/0803.1593} {\path{arXiv:0803.1593}},
  \href {http://dx.doi.org/10.1016/j.physrep.2008.10.001}
  {\path{doi:10.1016/j.physrep.2008.10.001}}.

\bibitem{Tytgat:1999yx}
M.~H. Tytgat, {QCD at theta similar to pi reexamined: Domain walls and
  spontaneous CP violation}, Phys.Rev. D61 (2000) 114009.
\newblock \href {http://arxiv.org/abs/hep-ph/9909532}
  {\path{arXiv:hep-ph/9909532}}, \href
  {http://dx.doi.org/10.1103/PhysRevD.61.114009}
  {\path{doi:10.1103/PhysRevD.61.114009}}.

\bibitem{Soto:1992vn}
J.~Soto, R.~Tzani, {Anomalies in the effective theory of heavy quarks},
  Phys.Lett. B297 (1992) 358--366.
\newblock \href {http://arxiv.org/abs/hep-ph/9210205}
  {\path{arXiv:hep-ph/9210205}}, \href
  {http://dx.doi.org/10.1016/0370-2693(92)91275-E}
  {\path{doi:10.1016/0370-2693(92)91275-E}}.

\bibitem{DiVecchia:1980ve}
P.~Di~Vecchia, G.~Veneziano, {Chiral Dynamics in the Large n Limit}, Nucl.Phys.
  B171 (1980) 253.
\newblock \href {http://dx.doi.org/10.1016/0550-3213(80)90370-3}
  {\path{doi:10.1016/0550-3213(80)90370-3}}.

\bibitem{Kawarabayashi:1980dp}
K.~Kawarabayashi, N.~Ohta, {The Problem of $\eta$ in the Large $N$ Limit:
  Effective Lagrangian Approach}, Nucl.Phys. B175 (1980) 477.
\newblock \href {http://dx.doi.org/10.1016/0550-3213(80)90024-3}
  {\path{doi:10.1016/0550-3213(80)90024-3}}.

\bibitem{Leutwyler:1992yt}
H.~Leutwyler, A.~V. Smilga, {Spectrum of Dirac operator and role of winding
  number in QCD}, Phys.Rev. D46 (1992) 5607--5632.
\newblock \href {http://dx.doi.org/10.1103/PhysRevD.46.5607}
  {\path{doi:10.1103/PhysRevD.46.5607}}.

\bibitem{Blaschke:2001ek}
D.~Blaschke, F.~Saradzhev, S.~Schmidt, D.~Vinnik, {A Kinetic approach to
  eta-prime production from a CP odd phase}, Phys.Rev. D65 (2002) 054039.
\newblock \href {http://arxiv.org/abs/nucl-th/0110022}
  {\path{arXiv:nucl-th/0110022}}, \href
  {http://dx.doi.org/10.1103/PhysRevD.65.054039}
  {\path{doi:10.1103/PhysRevD.65.054039}}.

\bibitem{Fukushima:2012fg}
K.~Fukushima, K.~Mameda, {Wess-Zumino-Witten action and photons from the Chiral
  Magnetic Effect}, Phys.Rev. D86 (2012) 071501.
\newblock \href {http://arxiv.org/abs/1206.3128} {\path{arXiv:1206.3128}},
  \href {http://dx.doi.org/10.1103/PhysRevD.86.071501}
  {\path{doi:10.1103/PhysRevD.86.071501}}.

\bibitem{Alles:1996nm}
B.~Alles, M.~D'Elia, A.~Di~Giacomo, {Topological susceptibility at zero and
  finite T in SU(3) Yang-Mills theory}, Nucl.Phys. B494 (1997) 281--292.
\newblock \href {http://arxiv.org/abs/hep-lat/9605013}
  {\path{arXiv:hep-lat/9605013}}, \href
  {http://dx.doi.org/10.1016/S0550-3213(97)00205-8}
  {\path{doi:10.1016/S0550-3213(97)00205-8}}.

\bibitem{Azcoiti:2003ai}
V.~Azcoiti, A.~Galante, V.~Laliena, {Theta vacuum: Phase transitions and / or
  symmetry breaking at theta = pi}, Prog.Theor.Phys. 109 (2003) 843--851.
\newblock \href {http://arxiv.org/abs/hep-th/0305065}
  {\path{arXiv:hep-th/0305065}}, \href {http://dx.doi.org/10.1143/PTP.109.843}
  {\path{doi:10.1143/PTP.109.843}}.

\bibitem{Aarts:2010gr}
G.~Aarts, K.~Splittorff, {Degenerate distributions in complex Langevin
  dynamics: one-dimensional QCD at finite chemical potential}, JHEP 1008 (2010)
  017.
\newblock \href {http://arxiv.org/abs/1006.0332} {\path{arXiv:1006.0332}},
  \href {http://dx.doi.org/10.1007/JHEP08(2010)017}
  {\path{doi:10.1007/JHEP08(2010)017}}.

\bibitem{Aarts:2013nja}
G.~Aarts, L.~Bongiovanni, E.~Seiler, D.~Sexty, I.-O. Stamatescu, {Complex
  Langevin simulation for QCD-like models}\href
  {http://arxiv.org/abs/1310.7412} {\path{arXiv:1310.7412}}.

\bibitem{Braaten:1991we}
E.~Braaten, M.~H. Thoma, {Energy loss of a heavy quark in the quark - gluon
  plasma}, Phys.Rev. D44 (1991) 2625--2630.
\newblock \href {http://dx.doi.org/10.1103/PhysRevD.44.2625}
  {\path{doi:10.1103/PhysRevD.44.2625}}.

\bibitem{Herzog:2006gh}
C.~Herzog, A.~Karch, P.~Kovtun, C.~Kozcaz, L.~Yaffe, {Energy loss of a heavy
  quark moving through N=4 supersymmetric Yang-Mills plasma}, JHEP 0607 (2006)
  013.
\newblock \href {http://arxiv.org/abs/hep-th/0605158}
  {\path{arXiv:hep-th/0605158}}, \href
  {http://dx.doi.org/10.1088/1126-6708/2006/07/013}
  {\path{doi:10.1088/1126-6708/2006/07/013}}.

\end{thebibliography}
\end{document}